\documentclass[preprint,showpacs,preprintnumbers,amsmath,amssymb]{revtex4}

\usepackage{graphicx}
\usepackage{epsfig}		
\usepackage{dcolumn}
\usepackage{bm}

\def\0{\mbox{\tiny $0$}}
\def\1{\mbox{\tiny $1$}}
\def\2{\mbox{\tiny $2$}}
\def\3{\mbox{\tiny $3$}}
\def\4{\mbox{\tiny $4$}}
\def\5{\mbox{\tiny $5$}}
\def\6{\mbox{\tiny $6$}}
\def\7{\mbox{\tiny $7$}}
\def\8{\mbox{\tiny $8$}}
\def\9{\mbox{\tiny $9$}}

\def\k{\mbox{\tiny $k$}}

\def\e{\mbox{\tiny $e$}}

\def\f14{\mbox{\tiny $\frac{1}{4}$}}

\def\L{\mbox{\tiny $L$}}
\def\V{\mbox{\tiny $V$}}

\def\ii{\mbox{\tiny $i$}}

\def\j{\mbox{\tiny $j$}}
\def\mi{\mbox{\tiny $-$}}

\def\bb#1{\mbox{\footnotesize $(#1)$}}

\begin{document}

\preprint{DF/IST-02.2008}
\preprint{February 2008}

\title{Lorentz violating extension of the Standard Model and the $\beta$-decay end-point}

\author{A. E. Bernardini}
\altaffiliation[On leave of absence from the]{~Instituto de F\'{\i}sica Gleb Wataghin, UNICAMP, PO Box 6165, 13083-970,
Campinas, SP, Brasil}
\email{alexeb@ifi.unicamp.br}
\author{O. Bertolami}
\altaffiliation[Also at]{~Instituto de Plasmas e Fus\~{a}o Nuclear}
\email{orfeu@cosmos.ist.utl.pt}
\affiliation{Instituto Superior T\'ecnico, Departamento de F\'{\i}sica, Av. Rovisco Pais, 1, 1049-001, Lisboa, Portugal}

\date{\today}

\begin{abstract}
The Standard Model extension with additional Lorentz violating terms allows for redefining the equation of motion
of a propagating left-handed fermionic particle.
The obtained Dirac-type equation can be embedded in a generalized Lorentz-invariance preserving-algebra through the
definition of Lorentz algebra-like generators with a light-like preferred axis.
The resulting modification to the fermionic equation of motion introduces some novel ingredients to the phenomenological
analysis of the cross section of the tritium $\beta$-decay.
Assuming lepton number conservation, our formalism provides a natural explanation for the tritium $\beta$-decay end-point
via an effective neutrino mass term without the need of a sterile right-handed state.
\end{abstract}

\pacs{03.30.+p, 11.30.Cp, 11.30.-j}
\keywords{Lorentz Violating Systems - Dispersion Relations - Dirac Equation}

\maketitle

\section{Introduction}

Although Lorentz symmetry is one of the most basic features of our description of nature, there has been evidence
in the context of string/M-theory \cite{Kos89,Kos95} and loop quantum gravity \cite{Gam99} that such a symmetry, at
least in principle, might be broken.
Observational information on the violation of Lorentz invariance would provide essential insights into the nature of the
fundamental theory of unification, however, no decisive experimental evidence has been detected so far.
Furthermore, the most recent results with regard to ultra-high energy protons suggest that there is no need for
violation of Lorentz invariance for explaining the data \cite{Aug07}.

However radical, the idea of dropping the Lorentz symmetry has been repeatedly considered in the literature.
For instance, a background or constant cosmological vector field has been suggested as a way to introduce a
velocity with respect to a universe's preferred frame of reference \cite{Phi66}.
It has also been proposed, based on the behaviour of the renormalization group $\beta$ function of non-Abelian gauge theories,
that Lorentz invariance could be just a low-energy symmetry \cite{Nie78}.
Furthermore, higher dimensional theories of gravity that are not locally Lorentz invariant have been considered
in order to obtain light fermions in chiral representations \cite{Wei84}.
The breaking of Lorentz symmetry due to nontrivial solutions of string field theory has been first discussed in Refs. \cite{Kos89,Kos95}.
These nontrivial solutions arise in the context of the string field theory of open strings and may have
striking implications at low energy.
The Lorentz violation could, for instance, give rise to the breaking of conformal symmetry and this together
with inflation may lie at the origin of the primordial magnetic fields which are required to explain the observed
galactic magnetic fields \cite{Ber99}.
In addition, putative violations of the Lorentz invariance could contribute to the breaking of CPT symmetry \cite{Kos95}.
Tensor-fermion-fermion interactions expected in the low-energy limit of string field theories give rise, in the
early universe, and after the breaking of CPT symmetry, to a chemical potential that creates in equilibrium a
baryon-antibaryon asymmetry in the presence of baryon number violating interactions \cite{Ber97B}.
In this scenario, the breaking of CPT symmetry allows for an explanation of the baryon asymmetry of the Universe \cite{Ber97B,Car06}.

These theoretical investigations have been considered in the context of a perturbative framework developed to
analyze certain classes of departures from Lorentz invariance.
Space-time translations along with exact rotational symmetry in the rest frame of the cosmic background radiation
have been, for instance, considered, also to treat small departures from boost invariance in this privileged frame \cite{Col99,Col97}.
Furthermore, inspired in the possibility of spontaneous symmetry breaking of Lorentz symmetry in string theory,
a Lorentz violating (LV) extension of the Standard Model (SM) has been developed \cite{Kos97}.
In this context, LV modifications to the Dirac equation and to the associated neutrino sector have become the
object of several phenomenological studies \cite{Ber07A,Ber07B,Bog04}.

Still from the theoretical point of view, the so-called {\em very special relativity} (VSR) approach is
based on the hypothesis that the space-time symmetry group of nature is smaller than the Poincar\'{e}
group, and consists of space-time translations described by only certain subgroups of the Lorentz group.
The formalism of VSR has been expanded for studying some peculiar aspects of neutrino physics with the
VSR subgroup chosen to be the 4-parameter group SIM(2) \cite{Gla06B}.
Since neutrinos are known to be massive, several mechanisms have been devised in order to allow for
neutrino masses in the Standard Model Lagrangian \cite{Bil03}.
An interesting implication of VSR is that it can endow neutrinos with an effective mass without the need of
violation of lepton number or additional sterile states \cite{Gla06B}.
In spite of not being Lorentz invariant, the lepton number conserving neutrino masses are VSR invariant.
There is, however, no certainty that neutrino masses have a VSR origin,
but if so, their magnitude may be an indication of the strength of the LV effects in other sectors.
For instance, a connection with the existence of a preferred axis in the cosmic radiation anisotropy might be examined.
This is particularly welcome as experimental evidence suggests that neutrinos are massive and this is incompatible with the SM structure.

Aiming to quantify LV effects in the neutrino sector, we consider the LV extension of the SM \cite{Kos97} and
follow the usual mathematical procedure for obtaining the corresponding dispersion relations and the equation of motion
for propagating left-handed fermionic particles \cite{Ber00} .
In particular, we compute the corrections to the dispersion relation arising from a LV extension of the SM and adapt
it in order to examine the neutrino sector.
From this LV SM extension, after combining boosts and rotations through a specific transformation, we introduce a
preferential direction with the aid of a light-like vector defined as $n_{\mu}$($\equiv(1,0,0,1)$), $n^{2}=0$.
The transformation is chosen to bring the equation of motion of left-handed neutrinos with a
dynamics similar to that of VSR in what concerns the existence of a preferred space direction, even though the
corresponding Lorentz algebra is preserved.
We find that this procedure gives origin for a neutrino effective mass effect without the need of a sterile right-handed state.
Interestingly, this effective mass term does affect the $\beta$-decay end-point.
Thus, the mechanism that we propose here introduces additional ingredients to the phenomenological
analysis of the tritium $\beta$-decay cross-section, which can be tested through modifications on its end-point.
The effects considered here are complementary to other studies of LV effects on other sectors of the
SM (see e. g. Ref. \cite{Kos08} for a complete list).

\section{LV extension of the SM to the neutrino sector}

It is widely believed that, in spite of its phenomenological success, the SM is most likely a low-energy
approximation of some more fundamental theory where unification with gravity is achieved and the hierarchy problem solved.
It is quite conceivable that, in the context of this more fundamental underlying theory, which is most
likely higher dimensional, CPT symmetry and Lorentz invariance may undergo spontaneous symmetry breaking \cite{Kos89,Kos95}.
If one assumes that this breaking extends down to the four-dimensional space-time, they might manifest themselves
within the SM and their effects detected. Notice also that in higher dimensional bulk-brane models,
it is possible that Lorentz invariance is spontaneously broken in the bulk space, but preserved on the brane,
as discussed in Ref. \cite{Ber06}.

In order to account for the CPT spontaneous breaking and LV effects, an extension to the minimal
$SU(3) \otimes SU(2) \otimes U(1)$ SM has been developed \cite{Kos97} based on the idea that CPT spontaneous
breaking and LV terms might arise from the interaction of tensor fields with Dirac fields once Lorentz tensors
acquire non-vanishing vacuum expectation values.
Interactions of this form are expected to arise, for instance, from the string field trilinear self-interaction,
as in the open string field theory \cite{Kos89,Kos95}.
In order to preserve power-counting renormalizability within the SM, only terms involving operators with mass
dimension four or less are considered.
The fermionic sector contains CPT-odd and CPT-even contributions to the extended Lagrangian density which,
including these LV terms, reads
\begin{eqnarray}
\mathcal{L}_{LV} &=&
\frac{1}{2}i \bar{\psi}\gamma_{\mu} \stackrel{\leftrightarrow}{\partial^{\mu}}\psi
+ a_{\mu} \bar{\psi}\gamma_{\mu}\psi
+ b_{\mu} \bar{\psi}\gamma_{\5}\gamma_{\mu}\psi
+\frac{1}{2}i c_{\mu\nu} \bar{\psi}\gamma^{\mu}\stackrel{\leftrightarrow}{\partial^{\nu}}\psi
\nonumber\\
&&~~~~~~~~~~~+\frac{1}{2}i d_{\mu\nu} \bar{\psi}\gamma_{\5}\gamma^{\mu}\stackrel{\leftrightarrow}{\partial^{\nu}}\psi
+ H_{\mu\nu} \bar{\psi}\sigma^{\mu\nu}\psi
- m \bar{\psi} \psi,
\label{gg1}
\end{eqnarray}
where the coupling coefficients $a_{\mu}$ and $b_{\mu}$ have dimensions of mass, $c_{\mu\nu}$ and $d_{\mu\nu}$
are dimensionless and can have both symmetric and antisymmetric components, while $H_{\mu\nu}$ has dimension of
mass and is antisymmetric.
All the LV coefficients are Hermitian and only kinetic terms are kept, since we are interested in deducing the
free particle energy-momentum relation.
These parameters are flavour-dependent and some of them may induce flavour changing neutral currents whether
non-diagonal in flavour.

In case of fermionic fields $\psi$ corresponding to purely chiral eigenstates with a negative (left-handed)
chiral quantum number, $\gamma_{\5}\nu = -\nu$, the mass dependent term and the $H_{\mu\nu}$ term
in the above Lagrangian density vanish.
In order to reduce the number of free parameters, the ones with dimension of mass ($a_{\mu}$ and $b_{\mu}$ )
and the dimensionless ($c_{\mu\nu}$ and $d_{\mu\nu}$) ones can be naturally regrouped
so that the effective LV Lagrangian density can be written as
\begin{equation}
\mathcal{L}_{LV} =
\frac{1}{2}i \bar{\nu}\gamma_{\mu} \stackrel{\leftrightarrow}{\partial^{\mu}}\nu
+ a_{\mu}\bar{\nu}\gamma_{\mu}\nu
+\frac{1}{2}i c_{\mu\nu} \bar{\nu}\gamma^{\mu}\stackrel{\leftrightarrow}{\partial^{\nu}}\nu.
\label{gg2}
\end{equation}
where, in order to simplify the notation, $b_{\mu}$ and $d_{\mu\nu}$ have been absorbed by $a_{\mu}$ and $c_{\mu\nu}$,
respectively, without any physical implication concerning the chirality of the particles.

Recall that in the {\em Dirac} picture, lepton number is conserved and neutrinos acquire their masses via Yukawa
couplings to sterile SU(2)-singlet neutrinos \cite{Kim93}.
In the {\em Majorana} picture, lepton number is violated and neutrino masses result from the seesaw mechanism involving
heavy sterile states or via dimension-6 operators resulting from {\em ad hoc} new interactions \cite{Moh86}.
As we shall see in the following, the lepton number conserving Lagrangian density (\ref{gg2}), for left-handed chiral
particles, suggests a generalization for the equation of motion.

Indeed, the Dirac-type equation of motion arising from Eq.~(\ref{gg2}),
\begin{equation}
\left[i\gamma^{\mu}\,\left(\partial_{\mu} + c_{\mu}^{\lambda}\,\partial_{\lambda}\right) + \gamma^{\mu} \,a_{\mu}\right] \nu_{\L} = 0,
\label{gg3}
\end{equation}
introduces a new quadratically invariant four-momentum $\tilde{p}_{\mu} = p_{\mu} + a_{\mu} + c_{\mu}^{\lambda}\,p_{\lambda}$
with an associated dispersion relation,
\begin{equation}
\tilde{p}_{\mu} \tilde{p}^{\mu} = p_{\mu} p^{\mu} + a_{\mu} a^{\mu} +p_{\lambda}p^{\beta} \, c^{\lambda}_{\mu} c^{\mu}_{\beta} + 2 ( a_{\mu} p^{\mu} + p_{\lambda} c^{\lambda}_{\mu} p^{\mu} + p_{\lambda} c^{\lambda}_{\mu} a^{\mu}) = 0.
\label{gg4}
\end{equation}
In the following, we examine the possibility of obtaining the above dispersion relation from a Lorentz
invariant framework, i.e. a setting which looks as if the Lorentz algebra holds.
For that, one must obtain a generator $D$ of a transformation $U\bb{p_{\mu},a_{\mu},c_{\mu\nu}}$
such that $U\bb{p_{\mu},a_{\mu},c_{\mu\nu}} \circ p_{\mu}  \equiv \tilde{p}_{\mu}\bb{p_{\mu},a_{\mu},c_{\mu\nu}}$.

Let us first define the momentum space $\mathcal{M}$, the four-dimensional vector space of momentum vectors, $p_{\mu}$.
In this space, the ordinary Lorentz generators act as
\begin{equation}
L_{\mu\nu} = p_{\mu}\tilde{\partial}_{\nu} - p_{\nu}\tilde{\partial}_{\mu},
\label{pp01}
\end{equation}
where $\tilde{\partial}_{\mu} \equiv \partial/\partial p^{\mu}$,
and we assume the Minkowski metric signature and that all generators are
anti-Hermitian (where our notation is as follows: $\mu,\,\nu = 0,\,1,\,2,\,3$, and $i,\,j,\,k = 1,\,2,\,3$ and $c = 1$).
The ordinary Lorentz algebra is constructed in terms of the usual rotations
$J^{\ii}\equiv \epsilon^{\ii\j\k}L_{\j\k}$ and boosts $K^{\ii} \equiv L^{\ii\0}$ as
\begin{equation}
[J^{\ii}, K^{\j}] = \epsilon^{\ii\j\k}K_{\k};~~~~[J^{\ii}, J^{\j}] = [K^{\ii}, K^{\j}] = \epsilon^{\ii\j\k}J_{\k}.
\label{pp02A}
\end{equation}
In order to introduce the non-linear action that modifies the ordinary Lorentz
generators, but that preserves its algebra, we suggest the following {\em Ansatz} for the generalized transformation,
\begin{equation}
D \equiv (a_{\nu} + p_{\beta}c^{\beta}_{\nu})\tilde{\partial}^{\nu},
\label{pp03}
\end{equation}
which acts on the momentum space as
\begin{equation}
D \circ p_{\mu} \equiv a_{\mu} + p_{\beta}c^{\beta}_{\mu}.
\label{pp04}
\end{equation}
Notice that the modified four-momentum $\tilde{p}_{\mu}$ does not arise from a conformal transformation.
Therefore, there is no general rule for obtaining the generator $D$ \cite{Ber08B}.
We assume that the new action can be considered to be a non-standard and non-linear embedding of the Lorentz
group into a modified non-conformal group which, despite the modifications, satisfies precisely the ordinary
Lorentz algebra (\ref{pp02A}).
To exponentiate the new action, we observe that
\begin{equation}
k^{\ii} = U^{^{\mi 1}}\hspace{-0.35 cm}\bb{D}\, K^{\ii}\, U\bb{D} ~~\mbox{and}
~~ j^{\ii} = U^{^{\mi 1}}\hspace{-0.35 cm}\bb{D} \,J^{\ii} \,U\bb{D},
\label{pp06}
\end{equation}
where the transformation $U\bb{D}$ for the LV-dependent term is given by $U\bb{D} \equiv{\exp[D]}$.
The non-linear representation is then generated by $U\bb{D}$ and, despite not being
unitary ($U\bb{D\bb{p_{\mu},a_{\mu},c_{\mu\nu}}} \circ p_{\mu} \neq p_{\mu}$), it must preserve
the algebra, which is enforced by the constraint
\begin{equation}
\left[[L_{\mu\nu},\,D\bb{p_{\mu},a_{\mu},c_{\mu\nu}}],\,D\bb{p_{\mu},a_{\mu},c_{\mu\nu}}\right] = 0,
\label{pp07}
\end{equation}
from which we can set
\begin{equation}
k^{\ii} = K^{\ii} + [K^{\ii}, \, D] ~~\mbox{and}~~ j^{\ii} =  J^{\ii} + [J^{\ii},\,D].
\label{pp06B}
\end{equation}
At this point, to explicitly constrain parameters $a_{\mu}$ and $c_{\mu\nu}$ so to satisfy the condition
Eq.~(\ref{pp07}), we compute the commutation relation
\begin{eqnarray}
[L_{\mu\nu},\,D] =[L_{\mu\nu},\, (a_{\beta} + p_{\lambda}c^{\lambda}_{\beta})\tilde{\partial}^{\beta}] &=&
(a_{\mu} \tilde{\partial}_{\nu} - a_{\nu}\tilde{\partial}_{\mu})\nonumber\\
&&~~~+ (p_{\mu} c_{\nu\beta} \tilde{\partial}^{\beta} - p_{\nu} c_{\mu\beta} \tilde{\partial}^{\beta} + p_{\lambda} c^{\lambda}_{\nu} \tilde{\partial}^{\mu} - p_{\lambda} c^{\lambda}_{\mu} \tilde{\partial}^{\nu}),
\label{pp08}
\end{eqnarray}
from which follows
\begin{eqnarray}
\left[[L_{\mu\nu},\,D],\, D\right] &=&
(a_{\lambda}c^{\lambda}_{\mu} \tilde{\partial}_{\nu} - a_{\lambda}c^{\lambda}_{\nu}\tilde{\partial}_{\mu}) + 2 p_{\lambda}(c^{\lambda}_{\nu}c_{\mu\alpha} - c^{\lambda}_{\mu}c_{\nu\alpha})\tilde{\partial}^{\alpha}\nonumber\\
&&~~ + (p_{\mu} c_{\nu\beta} c^{\beta}_{\alpha}\tilde{\partial}^{\alpha} - p_{\nu} c_{\mu\beta} c^{\beta}_{\alpha}\tilde{\partial}^{\alpha}
+ p_{\lambda} c^{\lambda}_{\beta} c^{\beta}_{\mu}\tilde{\partial}_{\nu} - p_{\lambda} c^{\lambda}_{\beta} c^{\beta}_{\nu}\tilde{\partial}_{\mu})
\label{pp09}.
\end{eqnarray}
If $c_{\mu\nu}$ is a symmetric tensor, $c_{\mu\nu} = 1/2 (q_{\mu}n_{\nu}+q_{\mu}n_{\nu})$,
then the second term in the above equation vanishes.
However,  in order to satisfy the condition Eq.~(\ref{pp07}), a stronger constraint must be set,
\begin{equation}
a_{\lambda}c^{\lambda}_{\mu} = c_{\nu\beta} c^{\beta}_{\alpha} = 0.
\label{pp10}
\end{equation}
This condition can be satisfied introducing a preferred direction with the help of a light-like vector
defined as $n_{\mu}\equiv(n_{\0},\mathbf{n})$, such that $c_{\mu\nu} = \alpha n_{\mu}n_{\nu}$
and $a_{\mu} = \mu s_{\mu}$  for $s_{\mu}n^{\mu} = 0$, that is, a light-like vector $s_{\mu} \equiv n_{\mu}$
or a space-like vector $s_{\mu} \equiv (0,\mathbf{s})$ with $\mathbf{n}\cdot\mathbf{s} = 0$.
Notice that the phenomenological coefficients $\mu$ and $\alpha$ have mass dimension one and zero, respectively.

The above constraints allow for obtaining a Lorentz-like algebra in terms of the generators given by Eq.~(\ref{pp06}) \cite{Ber07A}.
Therefore, for the chiral neutrino sector, the LV parameters modify the covariant momentum in a
way to allow for embedding it into a quasi-Lorentz invariance framework.
These transformations are not quadratically invariant in the momentum space.
However, there is a modified invariant $||U\bb{D\bb{p_{\mu},a_{\mu},c_{\mu\nu}}}\circ p_{\mu}||^{\2} = 0$
which leads to the following dispersion relation,
\begin{eqnarray}
||U\bb{D\bb{p_{\mu},a_{\mu},c_{\mu\nu}}}\circ p_{\mu}||^{\2} = p^{\2} + a^{\2} + 2 (a \cdot p) + 2\alpha (n \cdot p)^{\2} = 0
\label{pp13}
\end{eqnarray}
for which the $U$-invariance can be easily verified through application of the transformation $U\bb{D\bb{p_{\mu},a_{\mu},c_{\mu\nu}}}$.

Continuous deformations of Lie algebras have been extensively explored,
both from the mathematical and physical view points, in the context of Lie-algebra cohomology \cite{Lev67}.
Implications for doubly special relativity (DSR), for instance, have been considered in Refs. \cite{Ame01}.
Here we present a brief account so to allow for a simple and easy manipulation scheme for determining the
deformations of a given Lie algebra and its structure constants.

In this context, a similar procedure was performed for embedding VSR into a Lorentz preserving-algebra
framework \cite{Ber07B,Ber08B}, resulting in differences with respect to the original VSR formulation \cite{Gla06B},
for which space-time symmetries are subgroups of the Poincar\'e group.
These subgroups, characteristic of the VSR, contain space-time translations together
with at least a 2-parameter subgroup of the Lorentz group isomorphic to that generated by the association
of {\em boost} ($K$) and rotation ($J$) Lorentz generators, $K_x + J_y$ and $K_y - J_x$, which can
be embedded in a {\em quasi}-Lorentz algebra.
In here we have shown how a physical realization of the equation of motion derived from the LV SM Lagrangian
can be obtained from a deformed {\em quasi}-Lorentz algebra, in the same sense that the VSR physical realizations
can be re-obtained from a {\em quasi}-Lorentz algebra embedding \cite{Ber07B,Ber08B}.

Furthermore, there is one interesting and important consequence of the emergence of this {\em quasi}-Lorentz algebra.
As in the usual Lorentz algebra, this algebra can be interpreted both as the algebra of space-time
symmetries, the gauge algebra of gravity, and the algebra of charges associated to particles (energy-momentum and spin).
The idea of preserving the Lorenz-algebra in spite of modifying (i. e. deforming) the Lorentz generators
follows an analogous procedure as in DSR where a $\kappa$-deformed Poincar\'e (or Lorentz) algebra can
be interpreted as an algebra of Lorentz symmetries of momenta if the momentum space is a de Sitter space of curvature
$\kappa$ \cite{Gli01}.
In particular, it is suggestive that one can extend this algebra to the full phase space algebra of a point particle,
by adding four (non-commutative) coordinates \cite{Gli01} in the same way as it has been done for VSR \cite{Ber07B,Ber08B}.

Finally, in what concerns the phenomenological implications, it is important to emphasize that our results,
in spite of establishing a preferential direction, likewise in VSR, they lead to modified dispersion relations,
in opposition to what happens in that formalism.
This implies a fundamental difference in the calculation of cross sections.
However, as we shall see in the next section, the observable signals arising from
the SM extension are not significantly different from those of VSR or its {\em quasi}-Lorentz embedded version.

In what follows we shall disregard any effect related to flavour changing neutral currents when more than one
flavour is involved, and use the above dispersion relation to examine the
phenomenological implications to the tritium $\beta$-decay end-point.
We shall consider in our analysis the scenario where $a_{\mu}$ is also a light-like vector,
that is $a_{\mu} = \mu s_{\mu} = \mu n_{\mu}$, since, if it were space-like, LV effects would either
disappear or be phenomenologically unfeasible.

\section{Phase integral and the cross section of the $\beta$-decay}

Before analyzing the phenomenological implications of the new dispersion relation, let us first consider
the Lorentz invariant phase integral in the momentum space, $\tilde{p}$,
\begin{equation}
\int\frac{\mbox{d}^{\4}\tilde{p}}{(2\pi)^{\4}} 2\pi \delta(\tilde {p}^{\2}) \equiv \int\frac{\mbox{d}^{\4} p}{(2\pi)^{\4}} \, J\bb{\tilde {p}, p}\,2\pi \delta(\tilde {p}^{\2}\bb{p_{\mu},a_{\mu},c_{\mu\nu}}),
\label{pp14}
\end{equation}
where $J\bb{\tilde {p}, p}$ is the Jacobian determinant of the variable transformation $p \rightarrow \tilde{p}$,
\begin{equation}
J\bb{\tilde {p}, p} = \left|\frac{\partial \tilde{p}_{\beta}}{\partial p_{\lambda}}\right| = |\delta^{\lambda}_{\beta} + \alpha \, n^{\lambda}n_{\beta} | = 1 + \alpha \, n^{\mu}n_{\mu} = 1,
\label{pp15}
\end{equation}
given the constraint on $c_{\mu\nu}$.
For the purpose of computing cross sections involving neutrinos, it is convenient to write the phase integral in spherical coordinates as
\begin{equation}
\frac{1}{(2\pi)^{\3}}\int \mbox{d}\Omega \int\mbox{d}E\,\int \mbox{d}p\, p^{\2}  \delta(f\bb{p_{\mu},a_{\mu},c_{\mu\nu}}) = \int d\Omega \int\mbox{d}E\,\int \mbox{d}p\,  p^{\2} \frac{\delta(p)}{\left.(\partial f /\partial p)\right|_ {p = p\bb{E, \theta}}}
\label{pp16}
\end{equation}
where, from here onwards, $p \equiv |\mathbf{p}|$, $d\Omega = d(\cos{\bb{\theta}})\,d\varphi$, and
\begin{equation}
p\bb{E, \,\theta} \equiv p\bb{E, \,x}  = \frac{(\mu + 2\alpha E) x + \sqrt{(\mu^{\2}- 2 \alpha E^{^{\2}})x^{\2} + (1+ 2 \alpha) E^{^{\2}} + 2 \mu E}}{1 - 2 \alpha x^{\2}}
\label{pp17}
\end{equation}
is the root of the new dispersion relation,
\begin{equation}
f\bb{p_{\mu},a_{\mu},c_{\mu\nu}} \equiv
f\bb{p,\, E, \,\theta} \equiv
f\bb{p,\, E, \,x} =
p^{\2} - E^{^{\2}} - 2 \mu (E + p\, x) - 2 \alpha (E + p\, x)^{\2} = 0,
\label{pp18}
\end{equation}
and $x = -\cos{\bb{\theta}}$.
Upon integration in $\varphi$ and $p$ one obtains
\begin{equation}
\frac{1}{(2\pi)^{\2}}
\int^{^{+1}}_{_{-1}} \mbox{d} x
\int^{^{\infty}}_{_{0}} \mbox{d} E \frac{p^{\2}\bb{E, \,x}}{\left.(\partial f /\partial p)\right|_ {p = p\bb{E, x}}}
\label{pp19}
\end{equation}
where
\begin{equation}
\left.
\frac{\partial f}{\partial p} =
\right|_ {p = p\bb{E, x}}
= 2 \sqrt{(\mu^{\2}- 2 \alpha E^{^{\2}})x^{\2} + (1+ 2 \alpha) E^{^{\2}} + 2 \mu E}.
\label{pp20}
\end{equation}

Once we have established these new dynamical features, the analysis of the energy spectrum in the
end-point region of the tritium $\beta$-decay can be straightforwardly addressed.
This analysis corresponds actually to the well-known method of direct determination of the neutrino mass \cite{FerPer}.
The usual differential decay rate for the $d \rightarrow u\,e^{\mi}\,\bar{\nu}_{\e}$
transition is related to the decay amplitude by \cite{Hal84}
\small\begin{equation}
\mbox{d}\Gamma = G_{F}^{\2} \sum_{spins}\left|\bar{u}\bb{p_{\e}}\gamma^{\0} (1-\gamma_{\5})\upsilon\bb{p_{\nu}}\right|^{\2}
\frac{\mbox{d}^{\3}p_{\e}}{(2\pi)^{\3} E_{\e}}
\frac{\mbox{d}^{\3}p_{\nu}}{(2\pi)^{\3} E_{\nu}}
2\pi \delta\bb{E_{\0} - E_{\e} - E_{\nu}}
\label{pp30}
\end{equation}\normalsize
where $E_{\0}$ is the energy released to the lepton pair, $G_{F}$ is the Fermi constant, and the
indexes $e$ and $\nu$ refer to electron and neutrino variables, respectively.

For the new dynamics related to the dispersion relation Eq.~(\ref{pp18}), the phase space restriction is
modified by a change in the relevant matrix and in the phase integral, as quantified in Eq.~(\ref{pp19}).
Although the weak leptonic charged current $J^{\mu}$ must be modified to ensure its conservation,
the LV $\alpha$-dependent term contribution is entirely negligible near the end-point.
This yields a maximal correction of order $\mu/m_{\e}$, that is, this correction is suppressed by the electron mass.
Therefore, besides the modification to the neutrino phase integral, the other relevant contribution
arises from the square matrix element $\upsilon\bb{\tilde{p}}\bar{\upsilon}\bb{\tilde{p}}$,
\small\begin{equation}
\upsilon\bb{\tilde{p}}\bar{\upsilon}\bb{\tilde{p}} = \frac{1-\gamma_{\5}}{2} \tilde{p}^{\mu}\gamma_{\mu}
= \left[p^{\mu} \gamma_{\mu} + a^{\mu} \gamma_{\mu}+ p^{\beta}c_{\beta}^{\mu} \gamma_{\mu} \right],
\label{pp32}
\end{equation}\normalsize
where we have suppressed the neutrino index for simplicity.
Performing now the sum over spins, one obtains
\small\begin{equation}
\sum_{spins}\left|\bar{u}\bb{p_{\e}}\gamma^{\0} (1\mi\gamma_{\5})\upsilon\bb{\tilde{p}}\right|^{\2} =
8\left[E_{\e}\,\tilde{E} + \mathbf{p_{\e}}\cdot\tilde{\mathbf{p}}\right].
\label{pp33}
\end{equation}\normalsize
Notice that the element $\mathbf{p_{\e}}\cdot\tilde{\mathbf{p}}$ yields a null contribution after the angular
integration over ($\varphi_{\e},\,\theta_{\e}$) relative to the electron momentum coordinates.
Introducing the new neutrino phase integral Eq.~(\ref{pp19}), after rewriting Eq.~(\ref{pp30}) in terms of
Eq.~(\ref{pp33}) and performing the $\varphi_{\e},\,\theta_{\e}$ integration, the differential cross section
for the $\beta$- decay can be written as
\small\begin{eqnarray}
\frac{\mbox{d}\Gamma}{\mbox{d}p_{\e}} &=& p^{\2}_{\e}\frac{4 G^{\2}}{(2 \pi)^{\3}}
\int^{^{+1}}_{_{-1}} \mbox{d} x
\int^{^{\infty}}_{_{0}}\mbox{d} E \, \frac{p^{\2}\bb{E, \,x} \,\tilde{E}\bb{\bb{E, \,x}}}{\sqrt{(\mu^{\2}- 2 \alpha E^{^{\2}})x^{\2} + (1+ 2 \alpha) E^{^{\2}} + 2 \mu E}}
\delta\bb{E_{\0} - E_{\e} - E}
\label{pp35}
\end{eqnarray}\normalsize
where $\tilde{E}\bb{\bb{E, \,x}} = E + \alpha (E + p\bb{E, \,x}\, x)$.
Evaluating the integral over the $x$ and $E$ variables, one gets after some mathematical manipulation:
\small\begin{eqnarray}
\frac{1}{p^{\2}_{\e}}\frac{\mbox{d}\Gamma}{\mbox{d}p_{\e}} &=&\frac{G^{\2}}{\pi^{\3}}
\frac{1 - \alpha}{(1 - 2\alpha)^{\2}} (E + \mu)^{\2}
\label{pp36}
\end{eqnarray}\normalsize
where $E = E_{\0} - E_{\e} = (K_{max} + m_{\e}) - (K + m_{\e})$.

\begin{figure}
\centerline{\psfig{file=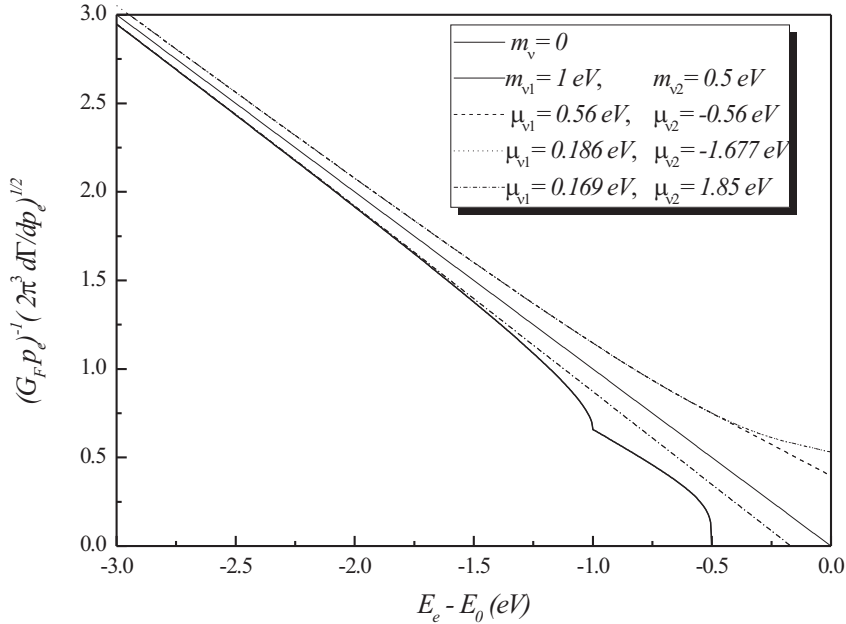,width=14.0 cm}}
\caption{The Kurie plot rate $p_{\e}^{\mi\1}(\mbox{d} \Gamma/ \mbox{d}p_{\e})$ as a function of the neutrino
energy ($E-E_{\0}$) near the end-point of the tritium $\beta$-decay spectrum for some set of LV parameters.
Values were chosen in order to qualitatively follow the asymptotic behaviour of the standard predictions for two
superimposed massive neutrino eigenstates with  $m_{\nu\1} = 1\, eV$ and $m_{\nu\2} = 0.5\, eV$.}
\label{fig1}
\end{figure}

At first glance, the Kurie plot rate $p_{\e}^{\mi\1}(\mbox{d} \Gamma/ \mbox{d}p_{\e})^{1/2}$
as a function of the neutrino energy ($E - E_{\0}$) near the end-point of the tritium $\beta$-decay
spectrum ($K_{\max} = 18.6\, $keV) for just one ({\em pseudo}) mass eingenstate does not seem to be phenomenologically interesting.
However, since the final state neutrinos are not detected in the tritium $\beta$-decays experiments,
for the electron spectrum, one should consider the incoherent sum
\small\begin{equation}
\frac{\mbox{d}\Gamma}{\mbox{d}p_{\e}} = \displaystyle\sum_{j = \1}^{\2}|U_{\e\j}|^{\2} \frac{\mbox{d}\Gamma}{\mbox{d}p_{\e}}\bb{\mu_{\j}, \alpha{\j}}.
\label{pp37}
\end{equation}\normalsize
In this case, by considering the possibility of superimposing, at least, two LV neutrino eigenstates,
$\nu\bb{\alpha_{\j},\mu_{\j}}$, $j = 1, \, 2$, one could easily reproduce the phenomenology of the
$\beta$-decay end-point for the usual neutrino mass scales if one imposes some constraints on the LV parameters.
In order to establish realistic values for the LV parameters $\alpha_{\j}$ and $\mu_{\j}$, we first
define the auxiliary phenomenological variables:
\small\begin{eqnarray}
a_{\1} \equiv \frac{1 - \alpha_{\1}}{(1 - 2\alpha_{\1})^{\2}} \sin{\bb{\theta}}^{\2}_{\bb{\L\V}},~~~
a_{\2} \equiv \frac{1 - \alpha_{\1}}{(1 - 2\alpha_{\1})^{\2}} \cos{\bb{\theta}}^{\2}_{\bb{\L\V}},
\label{pp38}
\end{eqnarray}\normalsize
subjected to the following constraints
\small\begin{eqnarray}
a_{\1} + a_{\2} ~=& 1& ~~\mbox{(Probability conservation)},\nonumber\\
a_{\1}\mu_{\1}  + a_{\2}\mu_{\2} ~=&  0 &~~\mbox{(Asymptotic behaviour)},\nonumber\\
a_{\1}\mu^{\2}_{\1} + a_{\2} \mu^{\2}_{\1} ~=& \frac{m^{\2}_{\1}+ m^{\2}_{\2}}{2}&~~\mbox{(Suitable order of magnitude)}.
\label{pp39}
\end{eqnarray}\normalsize
For typical values, say $m_{\1} = 1 \, eV$ and $m_{\2} = 0.5 \, eV$, we obtain the corresponding
LV parameters for some suitable choices of $a_{\1}$ and $a_{\2}$.
In Fig.~\ref{fig1} we compare the Kurie plot rate $p_{\e}^{\mi\1}\left(\mbox{d}\Gamma/\mbox{d}p_{\e}\right)^{\1/\2}$ with the usual ones.

We see that the tail of the spectrum is distinctly different for each preferred frame scenario.
The minimum of the curve corresponds to the case of massless neutrinos, so that, at the end-point, $K_{max} = E_{\0}= E_{\e}$.
For two of the three set of parameters that we have considered, one finds an excess (rather than a deficiency)
of events close to the end-point, as compared with the zero-mass case.
On quite general grounds, the knowledge of the neutrino mass spectrum is decisive for the understanding of the
origin of neutrino masses and mixing.
If, for instance, the KATRIN \cite{Katrin} experiment, currently in preparation, detects a positive effect
due to the neutrino mass, then $m_{\nu(\beta)} \approx m_{\nu_{\1,\2,\3}}$.
Whether non-vanishing neutrino mass effects are not observed, it is, of course, crucial to improve the
sensitivity of the $\beta$-decay experiments.
One should be aware that the KATRIN experiment, as well as its predecessors, measure the integrated
energy spectrum from the end-point downward.
This is proportional to
\small\begin{equation}
\Gamma\bb{K} = \int_{K}^{K_{max} - m_{\nu}}{\frac{\mbox{d}\Gamma}{\mbox{d}K}\mbox{d}K},
\label{pp40}
\end{equation}\normalsize
where $K(E_{\nu} = E - E_{\0})$ is defined as the electron kinetic energy $K = E - m_{\e} = K_{\max} + E_{\nu}$.
In any case, one can see that the proper knowledge of the experimental inputs allows for
fitting scenarios for values of $\alpha$ and $\mu$ and for comparison with the well-known mechanisms
for neutrino mass generation.
For sure, once experimental data are available, the effect of neutrino mass could be conveniently
expressed as the difference from the massless case in terms of $\Gamma_{m_{\nu}=\0}\bb{K} - \Gamma\bb{K}$
as a function of the neutrino energy ($E - E_{\0}$).

\section{Discussions and Conclusions}

In this work we have shown that the parameters of a LV extension of the SM have sizable implications for
the neutrino sector and, in particular, for the end-point of the tritium $\beta$-decay.
We have obtained a {\em non-conformal} transformation through which a new four-momentum is defined and hence a
new dispersion relation found. Although preserving the Lorentz algebra, we have implemented a preferred
direction scenario for the equation of motion of a propagating fermionic particle.
Focusing on the neutrino sector, the parameters of the LV extension of the SM can be directly confronted with the
next generation of tritium $\beta$-decay end-point experiments.

It is worth reminding that LV effects for the neutrino sector, concerning neutrino oscillation experiments and CPT violation,
were extensively studied in Ref. \cite{Mew04}.
The currently accepted solution for the oscillation data sets mass matrix elements in the $eV$-scale with mass-squared
differences of $10^{\mi\3}\,eV$ and $10^{\mi\5}\,eV$.
If one assumes that the mass matrix is nearly diagonal and that neutrino oscillations are primarily or entirely due to LV,
then individual masses of $\mathcal{O}(eV)$ or greater can be allowed with little or no effect on the existing
neutrino-oscillation data \cite{Mew04} even though, in the context of our analysis, we find that a non-negligible
signature in the $\beta$-decay end-point experiments is expected.

Actually, two other phenomenologically interesting scenarios are feasible: (i) Changes in the predictions
concerning neutrinoless double $\beta$-decay \cite{Ver02}, and (ii) Small changes in the
oscillation picture due to LV interactions that couple to active neutrinos, and which may
eventually allow for an explanation of all neutrino data \cite{Cor00}.
One could also mention that LV terms prevent the mechanism of Dirac chirality conversion \cite{Ber06D}
which is otherwise constrained for Dirac mass terms.
This could also alter phenomenological predictions concerning neutrino polarization \cite{Ber07D}.

Given that one of the most fundamental tasks in particle physics in the forthcoming future is the determination
of the neutrino mass scale, our proposal, which meets this end, can be also regarded, under conditions,
as a phenomenological implication of quantum gravity and string theory models. It is our believe that further
implications for cosmology and astrophysics which are worth being considered in the future.

\begin{acknowledgments}
A. E. B. would like to thank for the financial support from the FAPESP (Brazilian Agency) grant 07/53108-2 and the hospitality of the Physics Department of the Instituto Superior T\'{e}cnico, Lisboa, Portugal, where this work was carried out.
O. B. would like to acknowledge the partial support of Funda\c{c}\~ao para Ci\^encia e Tecnologia (Portuguese Agency)
under the project POCI/FIS/56093/2004.
\end{acknowledgments}

\end{document}